\documentclass[a4paper,11pt]{article}
\usepackage[T1]{fontenc}       
\usepackage[english]{babel}
\usepackage[latin1]{inputenc}
\usepackage{amsmath}
\usepackage{amsfonts}
\usepackage{amssymb}
\usepackage{amsthm}
\usepackage{dsfont}
\usepackage{graphicx}
\usepackage{natbib}
\usepackage{url}
\usepackage{rotating}
\usepackage[all]{xy}

\newcommand{\R}{\mathbb{R}}

\def\E{\mathds{E}}
\def\var{\mathrm{Var}}

\theoremstyle{plain}

\theoremstyle{definition}

\theoremstyle{remark}

\title{A Bayesian length-based population dynamics model for northern shrimp (\emph{Pandalus Borealis})}

\author{Paul Blomstedt\thanks{Fisheries and Environmental Management Group (FEM), Department of Environmental Sciences, P.O. Box 65 (Viikinkaari 2),
FIN-00014 University of Helsinki, Finland} \thanks{Helsinki Institute for Information Technology HIIT, Department of Computer Science, Aalto University}
\and Jarno Vanhatalo\footnotemark[1] 
\and Mats Ulmestrand\thanks{Department of Aquatic Resources, Swedish University of Agricultural Sciences, Sweden}
\and Anna G{\aa}rdmark\footnotemark[3]
\and Samu M\"antyniemi\footnotemark[1]
}

\date{}

\begin{document}
\maketitle

\begin{abstract}
We introduce a fully length-based Bayesian model for the population dynamics of northern shrimp (Pandalus Borealis). This has the advantage of structuring the population in terms of a directly observable quantity, requiring no indirect estimation of age distributions from measurements of size. The introduced model is intended as a simplistic prototype around which further developments and refinements can be built. As a case study, we use the model to analyze the population of Skagerrak and the Norwegian Deep in the years 1988--2012.
\end{abstract}

\section{Introduction}

Population dynamics models for shrimp generally fall into one of two categories, namely, surplus production models or age-structured models. In the former, the 
dynamics of a population are described simply in terms of rates of change of total biomass, while in the latter, age-dependent growth, growth-dependent recruitment to the fishery, and age-specific fishing and natural mortality are accounted for in great detail \citep{HvingelKingsley2006}. While it seems intuitive to follow the evolution of a population by age cohort, the determination of age is notoriously difficult in the case of shrimp \citep{QuinnEtAl1998} and  instead, age distributions have to be constructed indirectly by conversion from length. 
Despite this, very few length-based population dynamics models have been introduced for shrimp. This said, models do exist which make use of length-structured \emph{data}, while the population dynamics model itself is age-structured, see e.g. \citet{NielsenEtAl2013}.

In this paper, we present a model which is entirely length-based, such that no estimation of age distributions is required. The model structure is largely based on the general population dynamics model (GPDM) introduced in \citet{Mantyniemi_et_al_2015}, which is a state-space modeling framework for stock assessment introduced for enhanced biological realism in stock assessment. However, as the current model is a first attempt at a fully length-based model for shrimp, it has many simplifications compared to the full potential of GPDM.
The main focus of this paper is therefore on giving a description of this simplistic model in sufficient detail, so that it may serve as a prototype for further developments and refinements. 

This paper is structured as follows. In Sections \ref{sec:popdynmod} and \ref{sec:obs_mod}, we describe the population dynamics model and associated observation models in generic terms, and in Section \ref{sec:impl_res} we give details of implementation specific to Pandalus Borealis along with results of inference for the population of Skagerrak and the Norwegian Deep. Bayesian inference for the model is conducted using Markov Chain Monte Carlo (MCMC) as described in \citet{NewmanEtAl2009}. The paper concludes with a discussion in Section \ref{sec:discussion}.

\section{Population dynamics model}\label{sec:popdynmod}

Consider a population of interest which can be described in terms of a size distribution in numbers of individuals. 
The discretized distribution
\[
N_{t} = (N_{t,1},\ldots,N_{t,m}),
\]
will be referred to as the \emph{state} of the population at integer times $t=1,2,\dots$, where $N_{t,i}$ denotes the number of individuals belonging to size class $i=1,\ldots,m$ at at time $t$. A \emph{population dynamics model} is an idealized description of how the state of the population evolves from $N_{t-1}$ at time $t-1$ into $N_{t}$ at time $t$. Note, that the interval $(t-1,t]$ will be referred to as \emph{time period} $t$, so that for instance $(0,1]$ is time period 1 and $(1,2]$ is time period 2, whereas a given point in time, e.g. $t=1$ or $t=2$, will be referred to as \emph{time point} $t$, or for short, time $t$. 

\begin{figure}
\begin{displaymath}
\xymatrix{
												 & E_t \ar[rr]_{\text{ Recruitment }}     &  &N_t^R \ar@{=>}[dr]&\\
N_{t-1} \ar[ur]^{\text{ Reproduction }} \ar[dr]_{\text{ Growth }} &       & & &N_t\\ 
												 & N_t^G \ar[rr]^{\text{ Survival }}  \ar[drr] \ar[ddrr] & & N_t^S \ar@{=>}[ur]&\\
												 && &N_t^C&\\
												 && &N_t^D&}
\end{displaymath}
\caption{Schematic representation of the evolution of the population through time period $t$, 
indicating the key steps in the transition from $N_{t-1}$ to $N_t$.}
\label{fig:overview}
\end{figure}
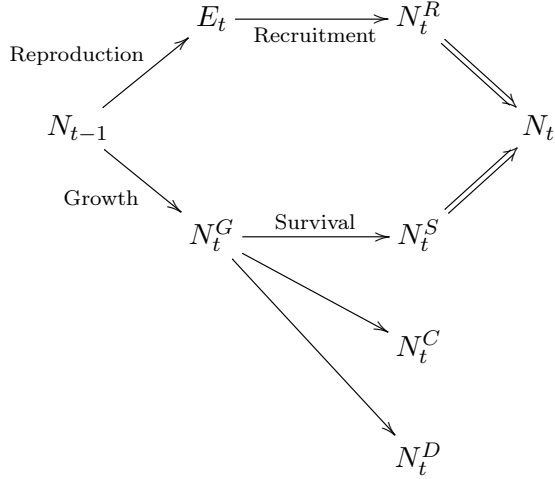

Throughout a time period $t$ there are various processes affecting the transition of the population from state $N_{t-1}$ into state $N_t$. While in reality, these processes act on the population gradually and simultaneously, they will for simplicity be modeled as being instantaneous and partly sequential. The main constituents of the population dynamics model are, on the one hand growth, survival and mortality, which affect the composition and size of the existing population, and on the other hand reproduction and recruitment, adding new individuals to the population. Figure \ref{fig:overview} gives an overview of the evolution of the population through time period $t$, where $N_t^G$ denotes the state of the population after growth, $N_t^S$, $N_t^C$ and $N_t^D$ denote the respective subpopulations pertaining to survival, fishing mortality and natural mortality, $E_t$ denotes the number of eggs produced in the beginning of time period $t$, and finally, $N_t^R$ denotes the population of recruits which enter the population at the end of time period $t$. The state of the population $N_t$ at time $t$ is then obtained as 
\[
N_t = N_t^S+N_t^R.
\]

In the following subsections, further details will be given on the specific models assumed for 
the different subprocesses of the population dynamics model. We begin with growth in Section \ref{sec:growth}, move on to survival and mortality in Section \ref{sec:survmort} and conclude with reproduction and recruitment in Section \ref{sec:reprod}. 
Throughout the remainder of this paper, we will denote the vector of breakpoints for the size classes by $I=(I_1,\ldots,I_{m+1})$, such that $I_i$ and $I_{i+1}$ are the lower and upper bounds, respectively, of size class $i$. The midpoint of the interval $[I_i,I_{i+1})$ corresponding to size class $i$ will be denoted by $l_i$.

\subsection{Growth}\label{sec:growth}

Growth is assumed to take place instantly in the beginning of each time period. A growth matrix which describes 
the movement of individuals between size classes is given by
\begin{equation}\label{eq:growthmatr}
\mathrm{\mathbf{G}}_t = \left(
\begin{array}{cccc}
g_{t,1,1} & g_{t,1,2} & \ldots & g_{t,1,m}\\
g_{t,2,1} & g_{t,2,2} & \ldots & g_{t,2,m}\\
\vdots & \vdots & \ddots & \vdots \\
g_{t,m,1} & g_{t,m,2} & \ldots & g_{t,m,m}
\end{array} \right), 
\end{equation}
where $g_{t,i,j}$ denotes the transition probability of moving from class $i = 1,\ldots,m$ to class $j=1,\ldots,m$, such that $\sum_{j=1}^m g_{t,i,j} = 1$ for all $i$. 

The probabilities $g_{t,i,j}$ are determined as follows. Denote by $L^{(i)}$ the length at time $t$ of individuals, which at time $t-1$ were in length class $i$. Assume then that $L^{(i)}$ is distributed as 
\begin{equation}\label{eq:distrL}
L^{(i)}\sim \mathrm{N}\Big(\mu^{(i)}_L,\sigma_L^2\Big)
\end{equation}
for all $t=1,2,\ldots$. The parameters of the distribution are obtained from the Von Bertalanffy (VB) growth model (Quinn and Deriso, 1999) \nocite{QuinnDeriso1999} as 
\begin{equation}\label{eq:muL}
\mu^{(i)}_L = (L_{\infty}-l_i)(1-e^{-k})+l_i 
\end{equation}
and
\begin{equation}\label{eq:sigma2L}
\sigma_L^2 = \sigma_{L_{\infty}}^2(1-e^{-2k}),        
\end{equation}
where $L_{\infty}$ and $\sigma_{L_{\infty}}^2$ are the asymptotic expected value and variance, respectively, for the length distribution of large individuals, and $k>0$ is a parameter controlling the growth rate.	Equation (\ref{eq:muL}) is a straightforward application of the VB growth model whereas (\ref{eq:sigma2L}) can be motivated by interpreting the VB model as an AR(1)-process, see Appendix \ref{app:VB} for details. Finally, denoting the cumulative distribution function of the normal distribution by $\Phi$, the probabilities $g_{t,i,j}$ are obtained from (\ref{eq:distrL}) as
\[
g_{t,i,j} = \frac{\Phi\Bigg(\frac{I_{j+1}-\mu^{(i)}_L}{\sigma_L}\Bigg)-\Phi\Bigg(\frac{I_j-\mu^{(i)}_L}{\sigma_L}\Bigg)}
{\Phi\Bigg(\frac{I_{m+1}-\mu^{(i)}_L}{\sigma_L}\Bigg)-\Phi\Bigg(\frac{I_1-\mu^{(i)}_L}{\sigma_L}\Bigg)}.
\]
Defined in this way, the transition matrix of Equation (\ref{eq:growthmatr}) is not strictly speaking an appropriate model for the growth of an individual, since there is a non-zero probability of moving to a lower size class. However, on a population level, it has been found to provide a more stable model than a triangular transition matrix, which has the undesirable effect of eventually accumulating the entire population into the highest size class \citep{Mantyniemi_et_al_2015}.

Assuming that individuals move between size classes independently, each row vector of $\mathrm{\mathbf{G}}_t$ corresponds to a probability vector for a multinomial distribution of 
$N_{t-1,i}$ independent trials,
\begin{equation}\label{eq:growth_multinom}
(n_{t,i,1}^G,n_{t,i,2}^G,\ldots,n_{t,i,m}^G)\sim\mathrm{Mult}\Big(N_{t-1,i}, (g_{t,i,1},g_{t,i,2},\ldots,g_{t,i,m})\Big),
\end{equation}
where $n_{t,i,j}$ denotes the number of individuals that moved from class $i$ to $j$. The state of the population after growth, $N_t^G$, is then found as the element-wise sum of all multinomially distributed  vectors,
\begin{align}\label{eq:NG}
N_t^G &= \Bigg(\sum_{i=1}^m n_{t,i,1}^G,\sum_{i=1}^m n_{t,i,2}^G,\ldots,\sum_{i=1}^m n_{t,i,m}^G\Bigg)\nonumber\\
			&= \big(N_{t,1}^G,\ldots,N_{t,m}^G\big).\phantom{\Bigg)}
\end{align}

Letting $N_{t-1,i}\rightarrow\infty$ in the distribution of Equation (\ref{eq:growth_multinom}) will shrink the relative standard deviation of each element $n_{t,i,j}^G$ towards zero. For large populations, this motivates approximating (\ref{eq:NG}) by its expected value
\begin{align}
N_t^G &\approx \E\big(N_t^G\big)\nonumber\\
      &= N_{t-1} \mathrm{\mathbf{G}}_t, \label{eq:NGapprox}
\end{align}
whereby considerable computational gains can be achieved.

\subsection{Survival and mortality}\label{sec:survmort}

Individuals in a given size class $i$ are assumed to have equal annual instantaneous natural and fishing mortality rates, denoted by $M_{t,i}$ and $F_{t,i},$ respectively. These rates are assumed to remain constant throughout each time period. In the current model, natural mortality is furthermore assumed to have a constant value $M_{t,i}=\gamma_M$ for all $t$ and $i$,
with $\gamma_M$ being treated as a static model parameter. The fishing mortalities of each length class are, on the other hand, assumed to be formed as class-specific fractions of a maximum attainable level of fishing mortality $F_t^\mathrm{max}$, which is assumed to vary by year, see further details in Section \ref{sec:fishmort} below. 

The total instantaneous mortality rate for time period $t$ is given by 
\[
Z_{t,i} =  M_{t,i} + F_{t,i}.
\]
Based on the instantaneous rates, the probabilities for individuals surviving to the next time period $\pi_{t,i}^S$, being caught by fishery $\pi_{t,i}^C$,  and dying naturally $\pi_{t,i}^D$,  can be found using the Baranov equations \citep[e.g.][]{Xiao2005}
\begin{align*}
\pi_{t,i}^S &= \exp(-Z_{t,i})\\
\pi_{t,i}^C &= \frac{F_{t,i}}{Z_{t,i}}\big(1-\exp(-Z_{t,i})\big)\\
\pi_{t,i}^D &= \frac{M_{t,i}}{Z_{t,i}}\big(1-\exp(-Z_{t,i})\big).
\end{align*}
The number of individuals pertaining to the different outcomes is modeled using a multinomial distribution 
\[
\big(N_{t,i}^S,N_{t,i}^C,N_{t,i}^D\big)\sim\mathrm{Mult}\Big(N_{t,i}^G,\big(\pi_{t,i}^S,\pi_{t,i}^C,\pi_{t,i}^D\big)\Big),
\]
where $N_{t,i}^G$ is the $i$:th element of vector (\ref{eq:NG}). By similar arguments as those which motivated the approximation of Equation (\ref{eq:NGapprox}), we may for large populations use the deterministic approximation
\[
\big(N_{t,i}^S,N_{t,i}^C,N_{t,i}^D\big) \approx \big(N_{t,i}^G\pi_{t,i}^S,N_{t,i}^G\pi_{t,i}^C,N_{t,i}^G\pi_{t,i}^D\big).
\]
Finally, the total numbers in the population are obtained as a sum of the class-wise numbers as
\[
\big(N_{t}^S,N_{t}^C,N_{t}^D\big) = \Bigg(\sum_{i=1}^m N_{t,i}^S,\sum_{i=1}^m N_{t,i}^C,\sum_{i=1}^m N_{t,i}^D\Bigg).
\]

\subsubsection{Fishing mortality}\label{sec:fishmort}

The marginal distribution of the maximum instantaneous fishing mortality $F_t^\mathrm{max}$ is assumed to be log-normal for all $t$,
\begin{equation*}\label{eq:Fmax_marginal}
F_t^\mathrm{max}\sim\mathrm{logN}(\mu_F,\sigma_F^2),
\end{equation*}
such that $\mu_F=\E\big(\log(F_1^\mathrm{max})\big)$ and $\sigma_F^2=\var\big(\log(F_1^\mathrm{max})\big)$. In practice, $\mu_F$ and $\sigma_F^2$ are expressed in terms of the parameters $\gamma_F$ and $CV_F$, such that $\mu_F=\log(\gamma_F)-\sigma^2_F/2$, where $\gamma_F=\E(F_t^\mathrm{max})$, and $\sigma_F^2=\log(CV_F^2+1)$, see Table \ref{tab:params}. In addition to the assumptions on the marginal distribution, we assume that the annual variation of $\log(F_t^\mathrm{max})$ around $\mu_F$ forms a dependent process over time. Therefore, we denote by $\xi_t$ the deviation of $\log(F_t^\mathrm{max})$ from $\mu_F$ in year $t$, so that 
\begin{equation}\label{eq:logFmax}
\log(F_t^\mathrm{max}) = \mu_F + \xi_{t},
\end{equation}
where
\begin{equation}\label{eq:xi_t}
\xi_t = \varphi_F\,\xi_{t-1} + \varepsilon_t
\end{equation}
is assumed to be a (wide-sense) stationary zero-mean AR(1)-process \citep{BoxEtAl2008} with $\varphi_F\in(0,1)$ and $\varepsilon_t\sim\mathrm{N}\big(0,(1-\varphi_F^2)\sigma_F^2\big)$.
From Equations (\ref{eq:logFmax}) and (\ref{eq:xi_t}) it then follows that	
\begin{align}\label{eq:Fmax_t}
F_t^\mathrm{max} &= \exp(\mu_F+\varphi_F\,\xi_{t-1}+\varepsilon_t) \nonumber\\
								 &= \exp\Big((1-\varphi_F)\mu_F+\varphi_F\big(\log(F_{t-1}^\mathrm{max})\big)+\varepsilon_t\Big).
\end{align}

Apart from annual variation, the instantaneous fishing mortality also depends on the size-dependent selectivity of the fishing gear \citep[e.g.][]{MillarFryer1999}, which for each class $i$ is obtained from a logistic function of length,
\begin{equation} \label{eq:selectivity}
f(l_i) = \frac{\exp(\alpha_f+\beta_f l_i)}{1+\exp(\alpha_f+\beta_f l_i)},\quad \alpha_f<0,\beta_f>0.
\end{equation}
We reparametrize the function in terms of $L_f^{(50)}=-\alpha_f/\beta_f$, which denotes the length at 50\% selectivity, and $\beta_f$, which controls the softness of the selection curve.
From Equations (\ref{eq:Fmax_t}) and (\ref{eq:selectivity}), the instantaneous fishing mortality of class $i$ is then finally obtained as 
\[
F_{t,i}  = f(l_i)\, F_{t}^\mathrm{max}.
\]

\subsection{Reproduction and recruitment} \label{sec:reprod}

The total number of recruits in time period $t$ is assumed be log-normally distributed, 
\[
R_t\sim\mathrm{logN}(\mu_{R_t},\sigma_R^2),
\]
where in particular, the parameter $\mu_{R_t}$ is related to the expected number of recruits $\E(R_t)$ through 
\[
\mu_{R_t} = \log\big(\E(R_t)\big)-\sigma_R^2/2.
\]
Also note, that $\sigma_R^2$ is in practice reparametrized as $\sigma_R^2=\log(CV_R^2+1)$, see Table \ref{tab:params}. 
It is now assumed that $\E(R_t)$ can be expressed in terms of a some function $h$ of the number of eggs $E_t$ produced in time period $t$, i.e.
\[
\E(R_t) = h(E_t).
\]
More specifically, assuming the functional form to be given by the Beverton-Holt stock-recruitment model, we have that  
\begin{equation}\label{eq:bh}
h(E_t) = \frac{K}{\frac{K}{\alpha}+E_t}E_t,
\end{equation}
where parameter $K:=\lim_{E_t\rightarrow \infty}h(E_t)>0$ has an interpretation as the asymptotic maximum expected number of recruits, and parameter $\alpha:=h'(0)\in(0,1)$, controls the rate at which $K$ is reached, see e.g. \citet{PulkkinenMantyniemi2013}. 

In Equation (\ref{eq:bh}), the total number of eggs produced in time period $t$ is calculated as  
\begin{equation} \label{eq:eggs}                                                    
E_t = \sum_{i=1}^m N_{t-1,i}\,e_{i}\,u_{i}\, \hat{r}(l_i),
\end{equation}
where $e_i$ denotes the proportion of females in size class $i$ and $u_i$ denotes the proportion of mature females among all females in class $i$, 
such that the proportion of mature females among all individuals in the size class is given by the product $e_i\, u_i$. The number of eggs $r$ produced by a female of length $l$ is commonly modeled using log-linear regression \citep[e.g.][]{ParsonsTucker1986}
\[
\log\big(r(l)\big) = \alpha_r^* + \beta_r\log(l)+\epsilon, \quad \alpha_r\in\R,\,\beta_r>0,
\]
where $\epsilon$ is an error term. Setting $\alpha_r^*=\log(\alpha_r)$, the average number of eggs produced by a mature female in size class $i$ is then obtained as 
\[
\hat{r}(l_i)=\alpha_r\,l_i^{\beta_r}.
\]

While the expected number of recruits $\E(R_t)$ is determined by the size of the population at time $t-1$ through Equations (\ref{eq:bh}) and (\ref{eq:eggs}), we will additionally assume that the random variation associated with $R_t$ has a dependency structure similar to that of fishing mortality in Section \ref{sec:fishmort}. To that end, let
\[
\zeta_t = \varphi_R \zeta_{t-1}+\upsilon_t,
\]
be a zero-mean AR(1)-process with $\varphi_R\in(0,1)$ and $\upsilon_t\sim\mathrm{N}\big(0,(1-\varphi_R^2)\sigma_R^2\big)$. The total number of recruits is now obtained as
\[
R_t = \exp(\mu_{R_t}+\varphi_R\,\zeta_{t-1}+\upsilon_t).
\]
In order to get the number of recruits $N_{t,i}^R$ per size class, we need to estimate the proportions $\phi_{i}^R$ of recruits entering each size class $i$. This is calculated as \[
\phi_{i}^R = \frac{\Phi\Bigg(\frac{I_{i+1}-\mu^{(0)}_L}{\sigma_L}\Bigg)-\Phi\Bigg(\frac{I_i-\mu^{(0)}_L}{\sigma_L}\Bigg)}
{\Phi\Bigg(\frac{I_{m+1}-\mu^{(0)}_L}{\sigma_L}\Bigg)-\Phi\Bigg(\frac{I_1-\mu^{(0)}_L}{\sigma_L}\Bigg)},
\]
where
\[
\mu^{(0)}_L = (L_{\infty}-L_0)(1-e^{-k})+L_0 
\]
is the expected length of recruits after one time period of growth, as given by the VB growth model. The length at age 0 is denoted by $L_0$ and is here for simplicity set to zero. The standard deviation $\sigma_L$ is obtained as the square root of Equation (\ref{eq:sigma2L}). The vector of recruits per size class in thus given as
\begin{align*}
N_t^R &= \big(R_t\phi_{1}^R,\ldots,R_t\phi_{m}^R\big)\\
      &= \big(N_{t,1}^R,\ldots,N_{t,m}^R\big).
\end{align*}

\section{Observation models}\label{sec:obs_mod} 

This section describes the statistical models which are used to link the population dynamics model described in Section \ref{sec:popdynmod} to observable data. The types of data used here are annual catch totals in tonnes and a survey biomass index. It is worth pointing out that the population dynamics model itself is not limited to these specific types of data. Thus, in general, any appropriate combination of models which probabilistically link available data to the population dynamics model could equally be applied.

\subsection{Total catch in tonnes}

The average weight (in grams) of an individual in length class $i$ is given by the relationship
\begin{equation*}\label{eq:length_weight}
w(l_i) = \alpha_w\, l_i^{\beta_w}.
\end{equation*}
Thus, the expected weight of the catch is
\[
\mu^U_t = \sum_{i=1}^m N^C_{t,i}w(l_i)/10^6,
\]
where the division by $10^6$ scales the weight from grams to tonnes. The observed weight of the catch $u_t$ in year $t$ is now assumed to be a realization of the random variable $U_t$,
\[
U_t\sim\mathrm{N}\big(\mu^U_t,\sigma^2_{\text{obs}}\big).
\]

\subsection{Survey biomass in tonnes}

The expected number of individuals per length class in the survey count is calculated as
\begin{align*}\label{eq:survey_n_est}
N_{t,i}^V &= q\,s(l_i)N_{t,i}^G\exp\big(-Z_{t,i}\delta_t^S\big) \nonumber\\
											&= q\,s(l_i)N_{t,i}^G\big(\pi_{t,i}^S\big)^{\delta_t^S},
\end{align*}
where $\delta_t^S$ denotes the relative time of the year of conducting the survey, $q$ is the catchability, which is a scaling factor between true biomass and biomass indices, and finally, 
$s(l_i)$ denotes the size-specific survey selectivity given by
\[
s(l_i) = \Phi\Bigg(\frac{l_i-L_s^{(50)}}{\sigma_s}\Bigg)\Bigg/\Phi\Bigg(\frac{l_m-L_s^{(50)}}{\sigma_s}\Bigg).
\]

Similarly to the observation model for catch, the expected survey biomass in tonnes is now given by
\[
\mu^V_t = \sum_{i=1}^m N^V_{t,i}w(l_i)/10^6.
\]
Assuming the observed survey biomass $v_t$ in year $t$ to be a realization of the random variable $V_t$, the observation model is then formulated as 
\[
V_t\sim\mathrm{N}\big(\mu^V_t,\sigma^2_{\text{obs}}\big).
\]

\section{Implementation and results} \label{sec:impl_res} 

We begin this section by giving some details about the implementation of the population dynamics model and the associated observation models described in the preceding sections. 
The state of the population is described in terms of a discretized length distribution, where the population is divided into $m=24$ length classes $[8,9)$, $[9,10)$, $\ldots$, $[31,32)$ of carapace length in units of 1 mm. Individuals outside this range are aggregated into classes at either endpoint of the range. The initial state of the population is given as 
\[
N_0 = \Bigg(N_{0,1}=\frac{K}{m},\ldots,N_{0,m}=\frac{K}{m}\Bigg),
\]
where the denominator of each element is here chosen to equal the carrying capacity $K$ of the Beverton-Holt model in Equation (\ref{eq:bh}). A summary of prior distributions for model parameters is given in Table \ref{tab:params}. While many of the prior distributions are entirely based on personal judgment, others are justified by existing knowledge found in the literature. In particular, two stock assessments conducted under the regime of the joint NAFO/ICES Pandalus Assessment Working Group and reported in \citet{Hvingel2013} and \citet{NielsenEtAl2013} give useful information for setting priors. Whenever a literary source has been used to provide justification for a prior distribution, either directly or in modified form, this has been indicated as a citation in Table \ref{tab:params}. For the fixed parameters $e_i$ and $u_i$ in Section \ref{sec:reprod}, we assume that $e_i\,u_i = 1$ for all $i$ such that $l_i>19$ and otherwise $e_i\,u_i = 0$.

\begin{sidewaystable} 
\centering
\begin{tabular}{lllllll}
\multicolumn{7}{l}{\textbf{Parameters for growth (Section \ref{sec:growth})}}\\
Parameter & \multicolumn{5}{l}{Prior distribution} & Reference\\
\hline
$L_{\infty}$ & $\mathrm{logitN}(\mu,\sigma^2;a,b)$ & $\mu=1,$&$\sigma^2 = 1,$&$a=26.5,$ &$b = 27.5$&\citet{NielsenEtAl2013}\\
$\sigma_{L_{\infty}}$ & $\mathrm{logitN}(\mu,\sigma^2;a,b)$&$\mu = 0,$ &$\sigma^2 = 1,$ &$a = 0.01,$ &$b = 1$&--\\
$k$ & $\mathrm{logitN}(\mu,\sigma^2;a,b)$&$\mu = 0,$ &$\sigma^2 = 1,$&$a = 0.4,$ &$b = 0.5$&\citet{NielsenEtAl2013}\\
\hline
&&&&&&\\
\multicolumn{7}{l}{\textbf{Parameters for survival and mortality (Section \ref{sec:survmort})}}\\
Parameter & \multicolumn{5}{l}{Prior distribution}& Reference\\
\hline
$\gamma_M$ & $\mathrm{logN}(\mu,\sigma^2)$&$\mu=\log(0.75),$&$\sigma^2=0.001$&&&\citet{NielsenEtAl2013}\\
$\gamma_F$ & $\mathrm{logN}(\mu,\sigma^2)$&$\mu=\log(0.4),$&$\sigma^2=0.05$&&&--\\
$CV_F$ & $\mathrm{Unif}(a,b)$&$a=0,$&$b=1$&&&--\\
$\varphi_F$ & $\mathrm{Unif}(a,b)$&$a=0.1,$&$b=0.9$&&&--\\
$L_f^{(50)}$ & $\mathrm{logN}(\mu,\sigma^2)$&$\mu=\log(18),$&$\sigma^2 = 0.005$&&&--\\
$\beta_f$ & $\mathrm{logN}(\mu,\sigma^2)$&$\mu=\log(0.3),$&$\sigma^2=0.1$&&&--\\
\hline
&&&&&&\\
\multicolumn{7}{l}{\textbf{Parameters for reproduction and recruitment (Section \ref{sec:reprod})}}\\
Parameter & \multicolumn{5}{l}{Prior distribution}& Reference\\
\hline
$\alpha_r$ & $\mathrm{logN}(\mu,\sigma^2)$&$\mu=-2,$&$\sigma^2=1$&&&\citet{ParsonsTucker1986}\\
$\beta_r$ & $\mathrm{N}(\mu,\sigma^2)$&$\mu=3,$&$\sigma^2=0.4$&&&\citet{ParsonsTucker1986}\\
$\alpha$ & $\mathrm{logitN}(\mu,\sigma^2)$&$\mu=-3,$&$\sigma^2=1$&&&--\\
$K$ &  $\mathrm{logN}(\mu,\sigma^2)$&      $\mu=23.025,$&$\sigma^2=1.33^2$&&&--\\
$CV_R$ & $\mathrm{logN}(\mu,\sigma^2)$&$\mu=\log(0.9),$&$\sigma^2=0.2$&&&--\\
$\varphi_R$ & $\mathrm{Unif}(a,b)$&$a=0.1,$&$b=0.9$&&&--\\
\hline
&&&&&&\\
\multicolumn{7}{l}{\textbf{Parameters for observation models (Section \ref{sec:obs_mod})}}\\
Parameter & \multicolumn{5}{l}{Prior distribution}& Reference\\
\hline
$\sigma_{\text{obs}}^2$ & $\text{Scale-inv-}\chi^2(\nu_0,\sigma^2_0)$&$\nu_0=15,$&$\sigma^2_0=1000^2$&&&--\\
$\alpha_w$ & $\mathrm{logN}(\mu,\sigma^2)$&$\mu=\log(0.00048271),$&$\sigma^2=0.01$&&&\citet{Wieland2002}\\ 
$\beta_w$ & $\mathrm{N}(\mu,\sigma^2)$&$\mu=3.0576,$&$\sigma^2=0.001$&&&\citet{Wieland2002}\\
$L^{(50)}_s$ & $\mathrm{logN}(\mu,\sigma^2)$&$\mu=\log(18),$&$\sigma^2=0.005$&&&--\\
$\sigma^2_s$ & $\text{Scale-inv-}\chi^2(\nu_0,\sigma^2_0)$&$\nu_0=10,$&$\sigma^2_0=50^2$&&&--\\
$q_1, q_3, q_4$ & $\mathrm{logN}(\mu,\sigma^2)$&$\mu=\log(0.173),$&$\sigma^2=0.3$&&&\citet{Hvingel2013}\\
\hline
\end{tabular}
\caption{Model parameters and their prior distributions. The notation $\mathrm{logitN}(\mu,\sigma^2;a,b)$ is used to denote a logit-normal distribution scaled to the interval $[a,b]$.
A reference has been cited whenever it has been used to provide either direct or indirect justification for a prior distribution.}
\label{tab:params}
\end{sidewaystable}

The data used for inference are annual catch totals in tonnes for the years 1988--2012 and a survey biomass index for the years 1984--2013, consisting of four different surveys, conducted in 1984--2002, 2003, 2004--2005 and 2006 onwards, see \citet{SovikThangstad2013} for further details. While the different surveys are not directly comparable, the indices give information about trends within the same survey. Since the second survey only covers the year 2003, providing no information about trends, it has been excluded. In Table \ref{tab:params}, the survey-specific catchability parameters for the included surveys are denoted $q_1$, $q_3$ and $q_4$. Inference was conducted using an implementation of a random-walk Metropolis-Hastings algorithm \citep[see e.g.][]{RobertCasella2004}, making use of the state-space model structure as outlined in \citet{NewmanEtAl2009}. After an initial tuning phase, the algorithm was run for 1.1 million iterations, from which the first 100\,000 iterations were discarded as burn-in. Finally, from the remaining 1 million iterations, every 100th sample point was retained to yield the final sample of size 10\,000. 

\begin{figure}
\centering
    \includegraphics[width=.85\textwidth]{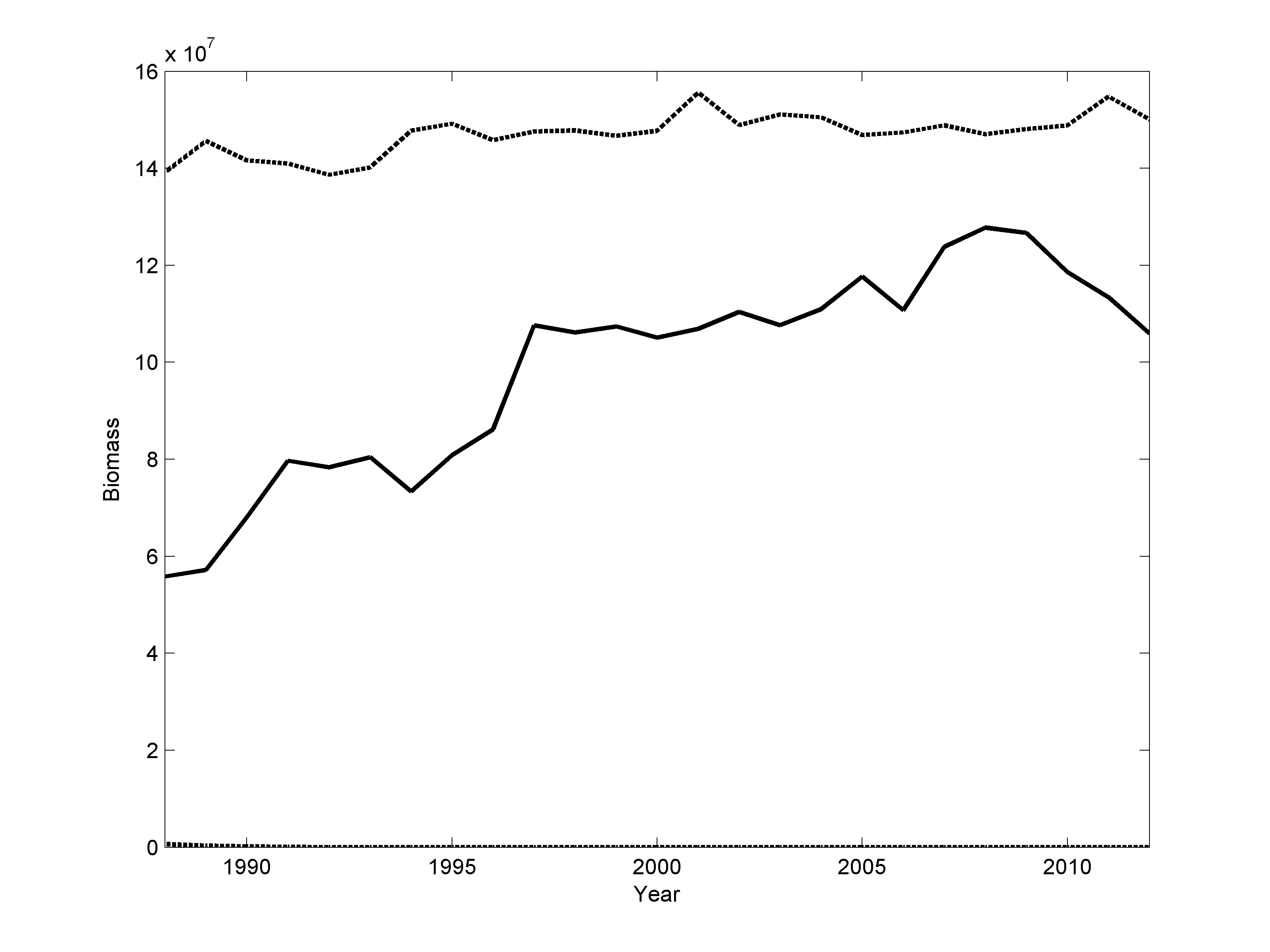}
    \caption{Mean estimates of total biomass in tonnes (solid) with 5\% and 95\% quantile envelopes (dashed).}
\label{fig:biomass1}
\end{figure}

\begin{figure}
\centering
    \includegraphics[width=.85\textwidth]{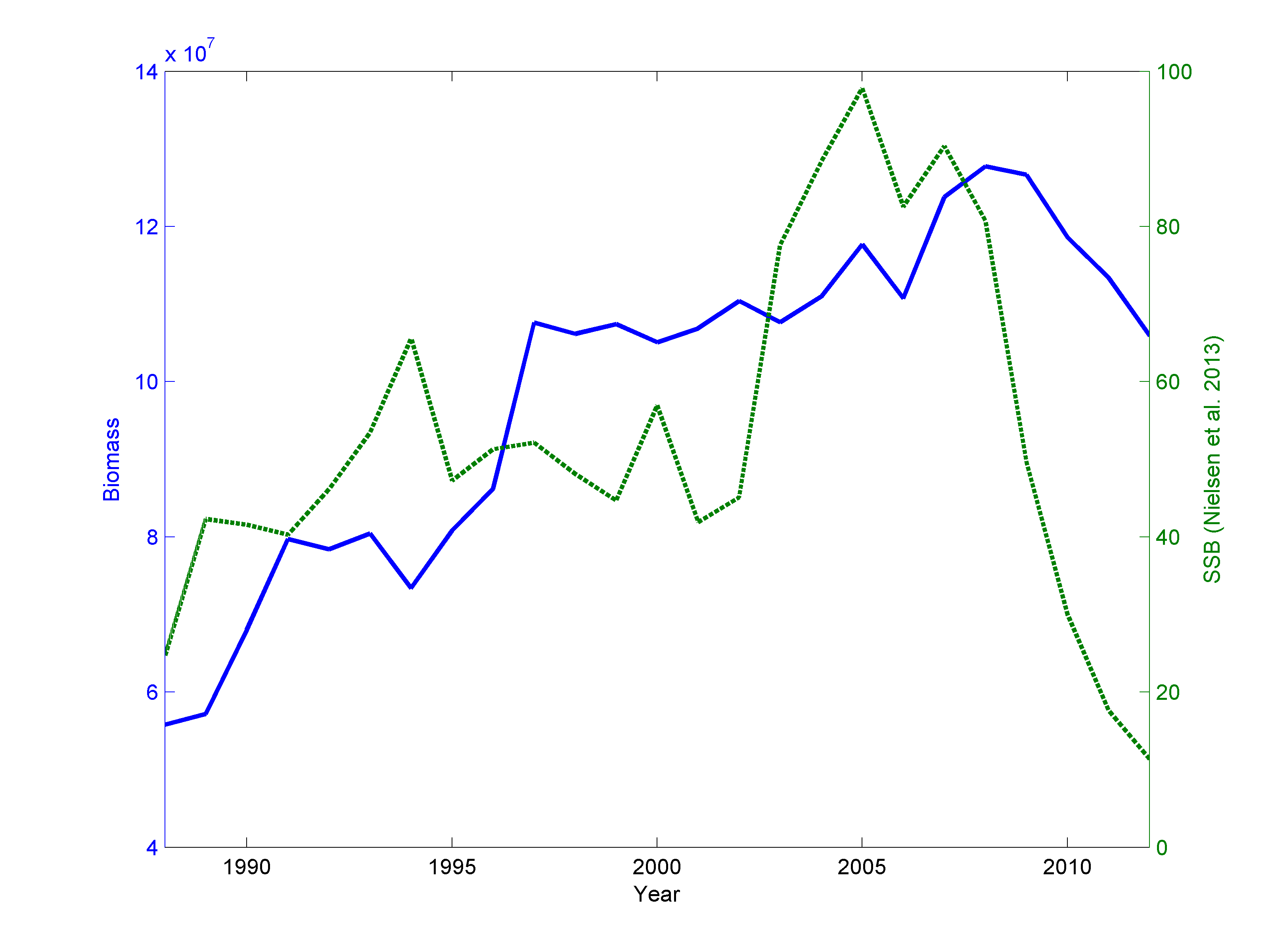}
    \caption{Comparison with biomass estimates of \citet{NielsenEtAl2013}.}
\label{fig:biomass2}
\end{figure}

\begin{figure}
\centering
    \includegraphics[width=.85\textwidth]{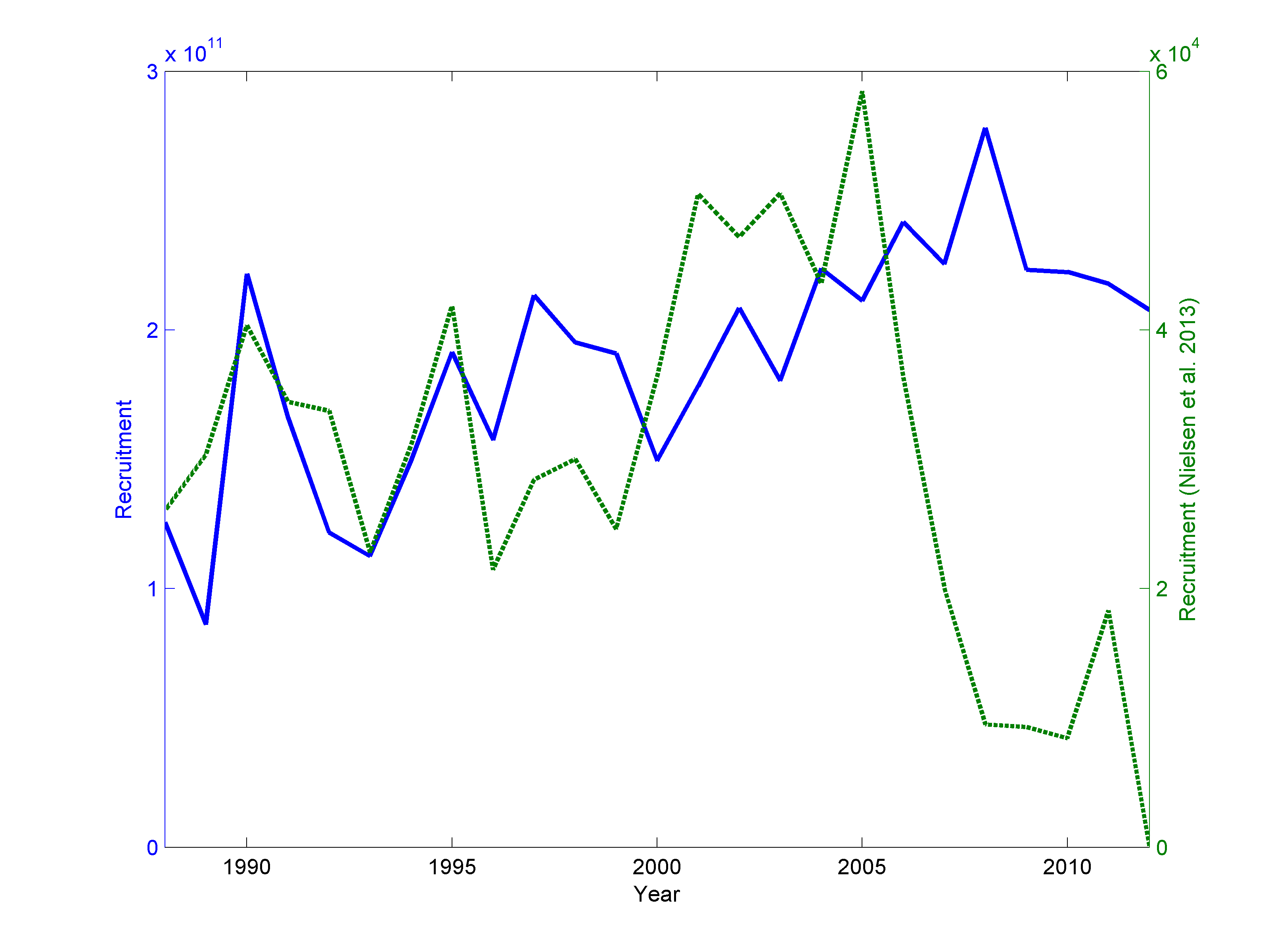}
    \caption{Recruitment in numbers of individuals compared with recruitment estimates of \citet{NielsenEtAl2013}.}
\label{fig:recruitment}
\end{figure}

\begin{figure}
\centering
    \includegraphics[width=.85\textwidth]{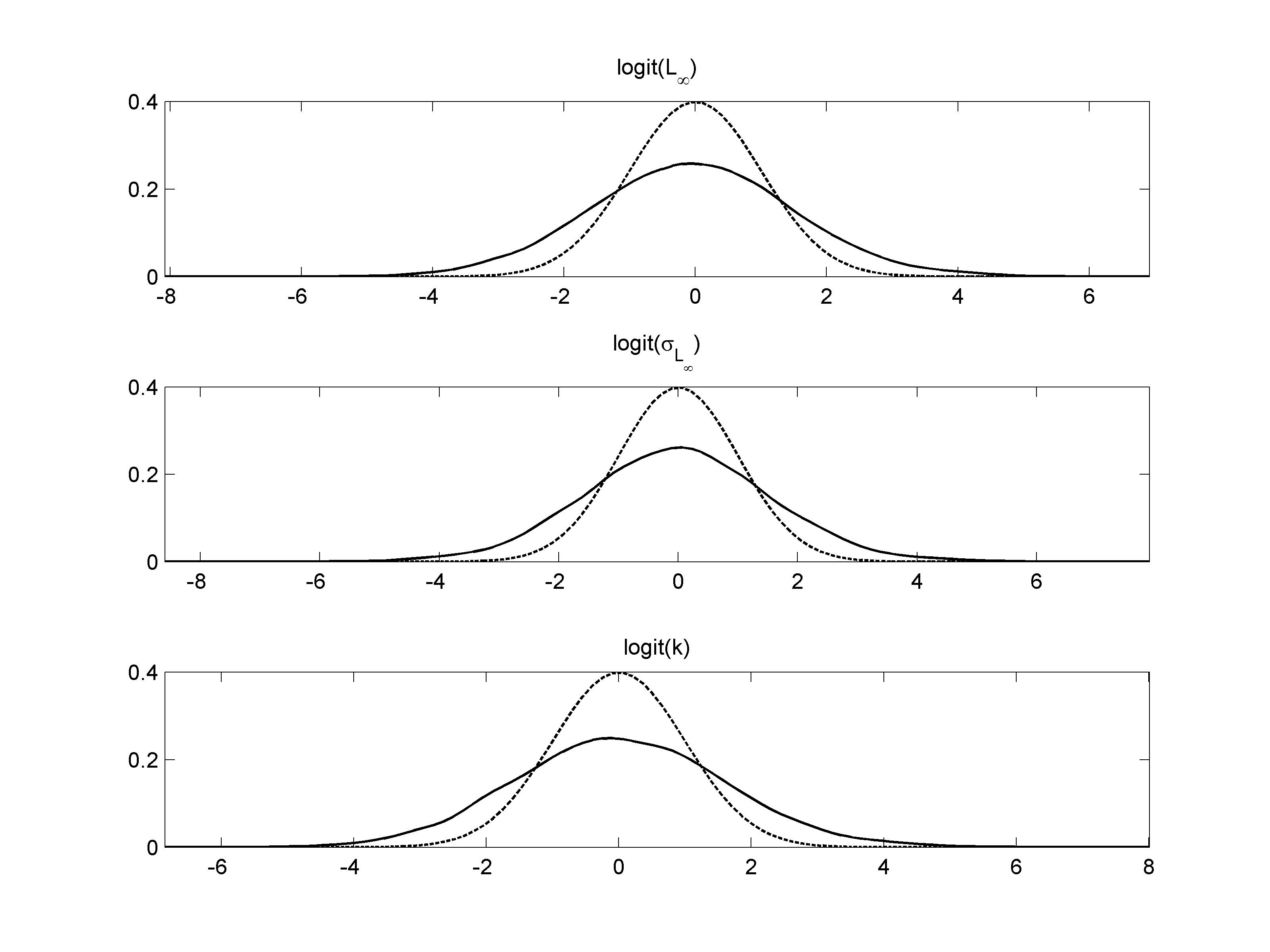}
    \caption{Posterior (solid) and prior (dashed) distributions for growth parameters.}
\label{fig:params1}
\end{figure}

\begin{figure}
\centering
    \includegraphics[width=.85\textwidth]{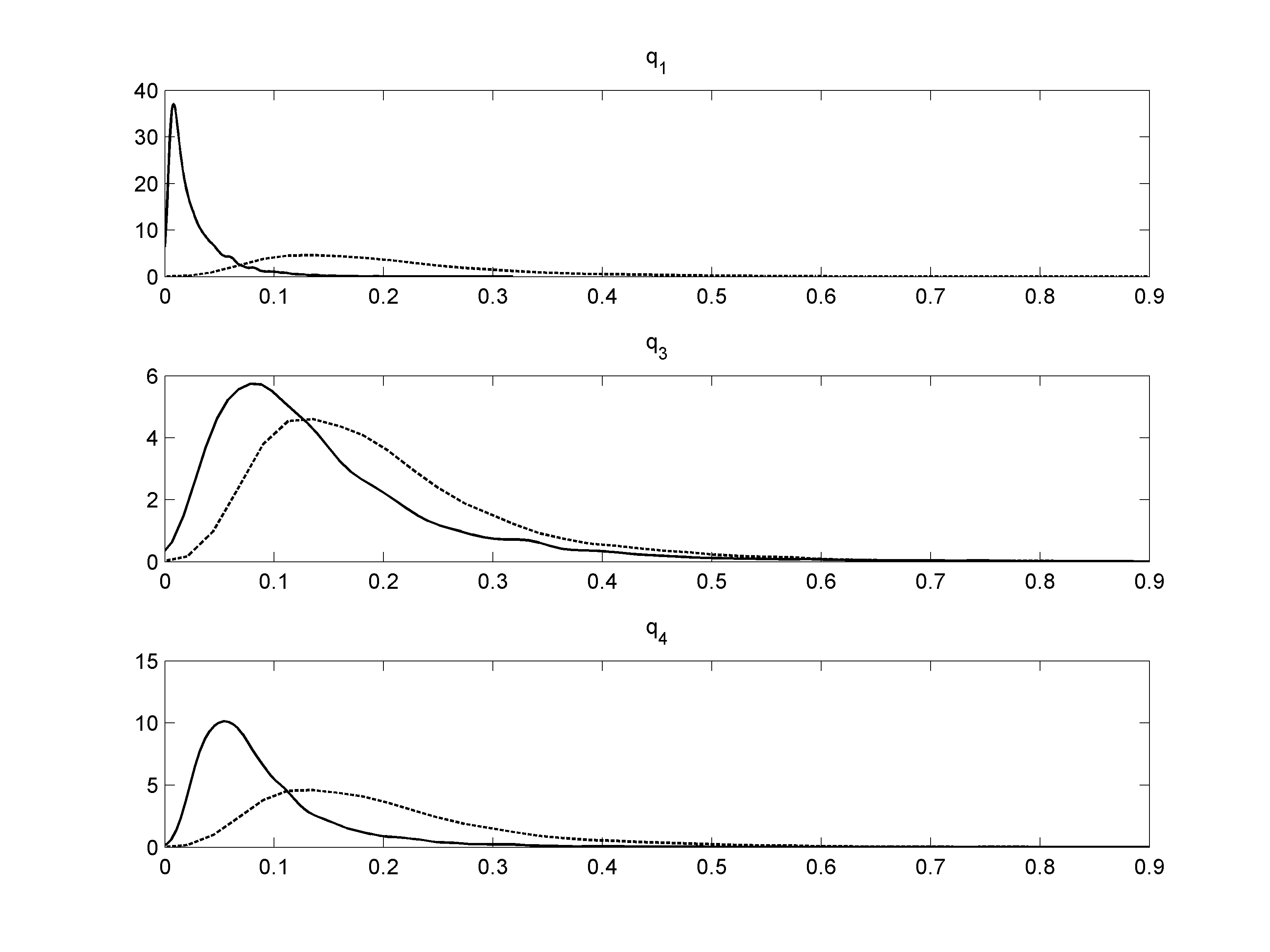}
    \caption{Posterior (solid) and prior (dashed) distributions for catchability in observation models.}
\label{fig:params5}
\end{figure}

The results of the inference are summarized for the years 1988--2012 in Figures \ref{fig:biomass1}--\ref{fig:params5}. These include estimates of total biomass calculated as $\sum_{i=1}^m N_{t,i}w(l_i)/10^6$, maximum fishing mortality $F_{t}^\mathrm{max}$, total number of recruited individuals $\sum_{i=1}^m N_{t,i}^R$, as well as posterior distributions for model parameters (note, that some of these distributions are shown for transformed parameters in order to enhance visual clarity). Figures \ref{fig:biomass2}--\ref{fig:recruitment} show a comparison with results reported for the stock assessment model of \citet{NielsenEtAl2013}. In particular, it is seen that the point estimates for the main outcome of total biomass show a similar increasing trend for both models which then turns into a decline between the years 2005--2010. The considerable uncertainty around this estimate revealed by Figure \ref{fig:biomass1} is, however, worth noting.

\begin{figure}
\centering
    \includegraphics[width=.85\textwidth]{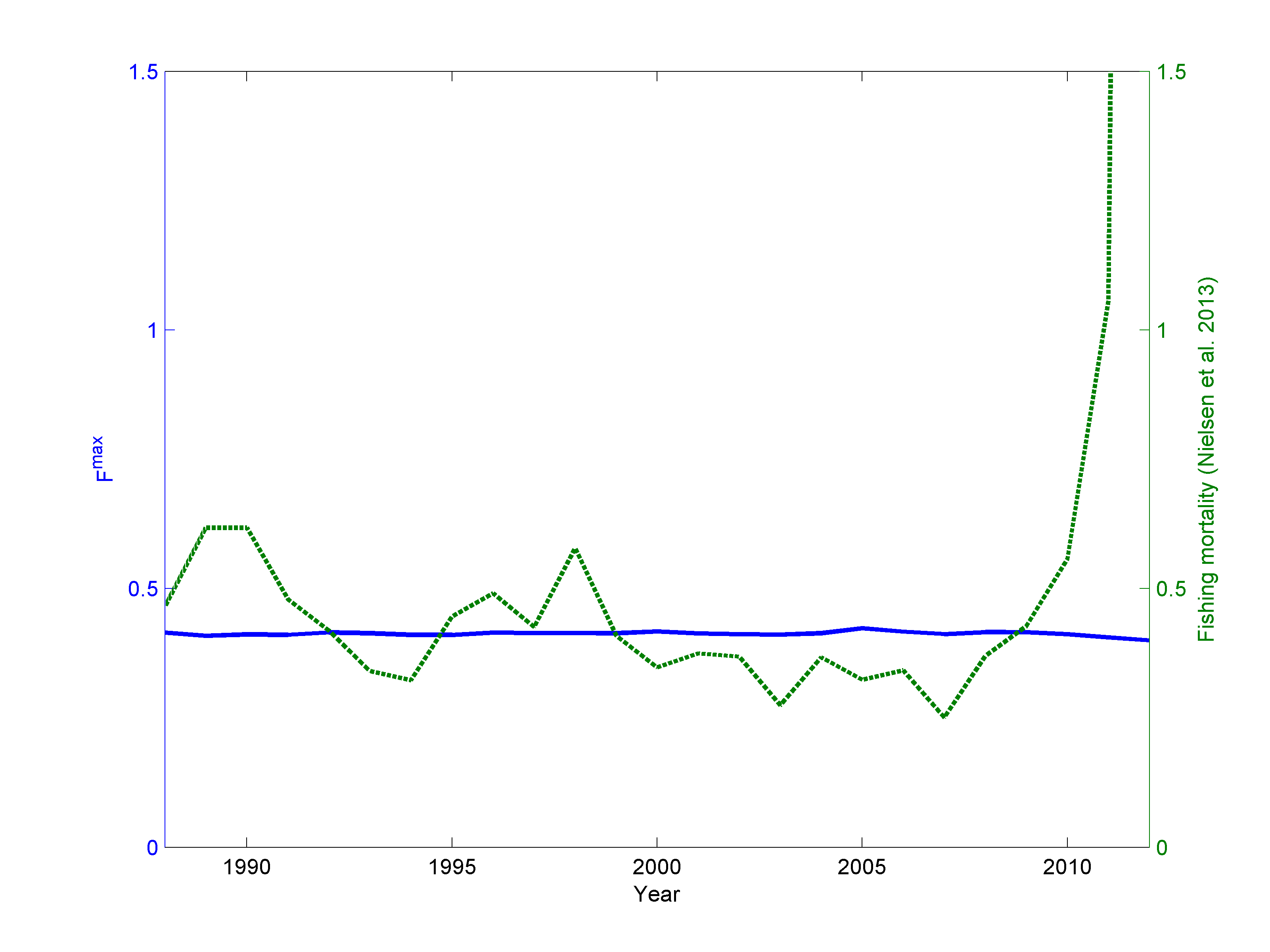}
    \caption{Maximum instantaneous fishing mortality compared with fishing mortality estimates of \citet{NielsenEtAl2013}.}
\end{figure}

\begin{figure}
\centering
    \includegraphics[width=.85\textwidth]{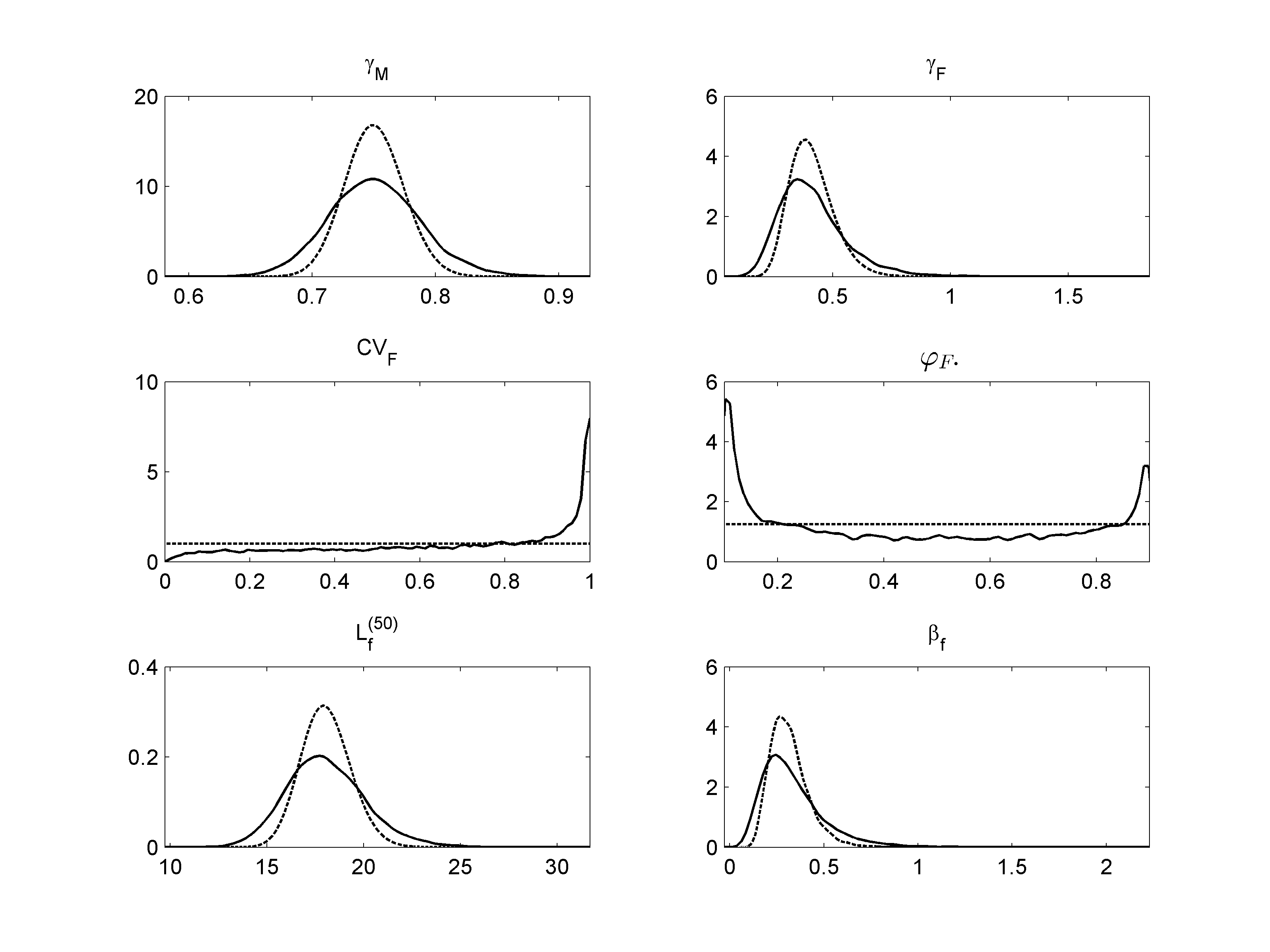}
    \caption{Posterior (solid) and prior (dashed) distributions for mortality parameters.}
\end{figure}

\begin{figure}
\centering
    \includegraphics[width=.85\textwidth]{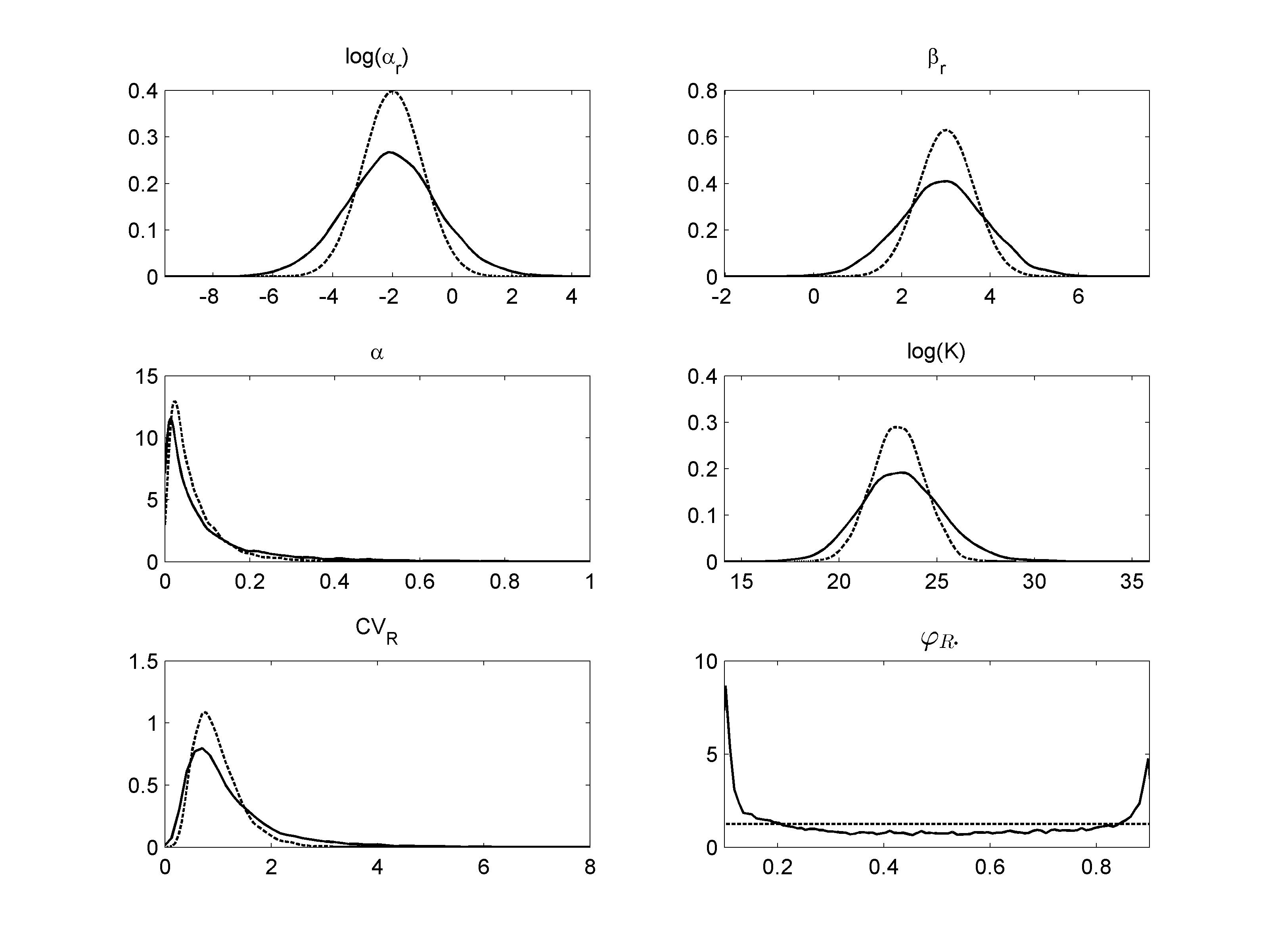}
    \caption{Posterior (solid) and prior (dashed) distributions for reproduction and recruitment parameters.}
\end{figure}

\begin{figure}
\centering
    \includegraphics[width=.85\textwidth]{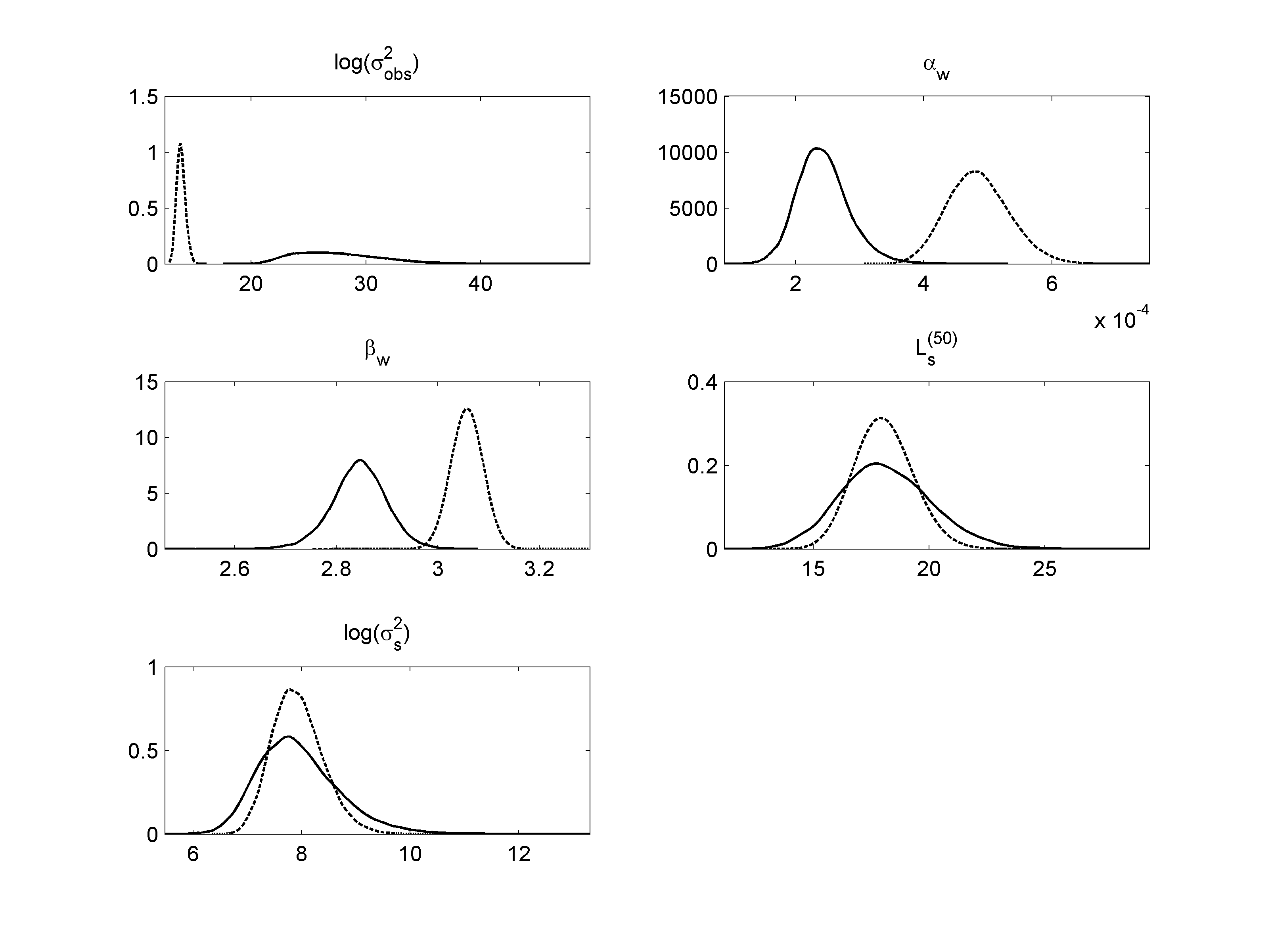}
    \caption{Posterior (solid) and prior (dashed) distributions for observation model parameters.}
\end{figure}

\section{Discussion} \label{sec:discussion}

In this paper, we have presented a fully length-based model for the population dynamics of northern shrimp (Pandalus Borealis). In comparison with age-based models, this has the advantage of structuring the population in terms of a directly observable quantity, requiring no indirect estimation of age distributions from size \citep[e.g.][]{MacdonaldPitcher1979}. The introduced model is a simplistic implementation of the general population dynamics model of \citet{Mantyniemi_et_al_2015} and intended as a prototype around which further developments can be built. These developments may involve more accurate modeling of biological and fisheries-induced processes, the use time steps shorter than the period of one year currently used, the inclusion of environmental covariates and additional process errors, as well as the utilization of more informative data. 

An aspect which to some extent currently limits the complexity of the model is the computation required for posterior inference. As the model grows increasingly complex, both the execution and convergence of the computational algorithm slows down to the point where inference on the model is impractical. Developing the model further should therefore entail the design of computational strategies more apt for state-space models than the currently used MCMC algorithm. An example of a class of simulation-based algorithms with great potential and which conform particularly well to these type of models are sequential Monte Carlo methods \citep[e.g.][]{Doucet_et_al_2001,Chopin_et_at_2013}. To that end, a simplistic prototype model, such as the current one, also serves the purpose of being a platform for testing novel computational ideas.

\section*{Acknowledgements}
The authors wish to thank Prof. Henrik Gislason at the Technical University of Denmark for a helpful discussion during the completion of the manuscript.

\bibliographystyle{natbib}
\bibliography{Pandalus_references}

\appendix

\section*{Appendix}

\section{Derivation of Equation (\ref{eq:sigma2L})} \label{app:VB}

Consider the Von Bertalanffy growth model
\begin{equation} \label{eq:VB}
L_{\tau} = (L_{\infty}-L_{\tau-1})(1-e^{-k})+L_{\tau-1} +\varepsilon_{\tau} 
\end{equation}
with an additional error term $\varepsilon_{\tau}\sim\mathrm{N}(0,\sigma_L^2)$, for all $\tau=1,2,\ldots$. 
Here, the time index $\tau$ refers to the growth history of an individual, so that $L_{\tau}$ is the length of the individual at age $\tau$.
Assume now that $\tau>M$ for some constant $M$. By setting $c = (1-e^{-k})L_{\infty}$ and $\varphi = e^{-k}$, the model (\ref{eq:VB}) can then be interpreted as an AR(1)-process
\[
L_{\tau} = c+\varphi L_{\tau-1}+\varepsilon_{\tau},
\]
for which, by the assumptions of Section \ref{sec:growth}, $\E(L_{\tau})=L_{\infty}$ and $\var(L_\tau) = \sigma_{L_{\infty}}^2$ as $M\rightarrow\infty$. Finally, by standard properties of AR(1)-processes it follows that 
\[
\sigma_L^2 = \sigma_{L_{\infty}}^2(1-e^{-2k}).
\]

\end{document}